\documentclass[aps, pra, reprint, lengthcheck, footinbib, showpacs,  citeautoscript]{revtex4-2}

\usepackage{bm}
\usepackage{graphicx}
\usepackage{amssymb}
\usepackage{amsmath}
\usepackage{eufrak}
\usepackage{color}
\usepackage[utf8]{inputenc}
\usepackage[unicode=true,colorlinks=true,citecolor=blue,urlcolor=blue]{hyperref}

\usepackage{ulem}

\renewcommand{\emph}{\textit}
\usepackage{braket}

\usepackage{chngcntr}

\graphicspath{{figs/}}

\newcommand{\nix}[1]{}

\newcommand{\cEK}[1]{\textcolor[rgb]{0,0.5,0}{{\bf Erik:} \emph{#1}}}

\begin{document}

\title{Resonant spin amplification and accumulation in MAPbI\texorpdfstring{$_3$}{3} single crystals}

\author{Erik Kirstein$^{1}$, Dmitri R. Yakovlev$^{1}$, Evgeny~A.~Zhukov$^{1}$,  Nataliia E. Kopteva$^{1}$, Bekir Turedi$^{2,3}$, Maksym V. Kovalenko$^{2,3}$, Manfred Bayer$^{1,4}$}

\affiliation{$^{1}$Experimentelle Physik 2, Technische Universit\"at Dortmund, 44227 Dortmund, Germany}
\affiliation{$^{2}$Department of Chemistry and Applied Biosciences,
Laboratory of Inorganic Chemistry, ETH Z\"{u}rich, 8093 Z\"{u}rich, Switzerland}
\affiliation{$^{3}$Department of Advanced Materials and
Surfaces, Laboratory for Thin Films and Photovoltaics, Empa - Swiss Federal Laboratories for Materials Science and Technology, 8600 D\"{u}bendorf, Switzerland}
\affiliation{$^{4}$Research Center FEMS, Technische Universität Dortmund, 44227 Dortmund, Germany}

\begin{abstract}
Quantum technologic and spintronic applications require reliable semiconducting materials that enable a significant, long-living spin polarization of electronic excitations and offer the ability to manipulate it optically in an external field.  Due to the specifics of band structure and remarkable spin-dependent properties, the lead halide perovskite semiconductors are suitable candidates for that. Here, the carrier spin dynamics in a MAPbI$_3$ (MA = methylammonium) perovskite single crystal with thickness of 20~$\mu$m are studied by the time-resolved Kerr ellipticity technique at cryogenic temperatures. Long times of longitudinal electron spin relaxation $T_1 = 30$\,ns and transverse electron spin dephasing $T_{2,e}^*=21$\,ns are found. The spin dynamics lasting longer than the applied laser pulse repetition period give rise to spin accumulation effects. We exploit them through the resonant spin amplification, polarization recovery, and spin inertia techniques to study the electron and hole spin systems coupled with the nuclear spins. These results establish the lead halide perovskite semiconductors as suitable platform for quantum technologies relying on spin-dependent phenomena.
\end{abstract}

\maketitle

\section{Introduction}
\label{Intro}
Lead halide perovskite semiconductors have evolved towards a versatile platform for photovoltaic and optoelectronic applications~\cite{vardeny2022,vinattieri2021,Martinez2023_book}. Their spin-dependent properties are promising for spintronic and spin-orbitronic applications~\cite{vardeny2022,privitera2021,wang2019,kim2021}. Band structure and crystal symmetry of the lead halide perovskites differ considerably from that of conventional III-V and II-VI semiconductors with zinc blende structure, establishing them as novel material platform for the spin physics of semiconductors~\cite{kirstein2022uni,kopteva2024OO}. Spin-dependent parameters, like the Land\'e $g$-factor, and spin relaxation mechanisms of electrons, holes and excitons, as well as nuclear spin dynamics provide access to the band parameters and their anisotropy, which often cannot be measured by other techniques. In turn, knowledge of these parameters, combined with flexibility of synthesis of the lead halide perovskites with band gaps tunable across the whole visible spectral range from 1.5 to 3.2~eV, allows tailoring of the spin-dependent properties.  

Time-resolved Faraday/Kerr rotation are widely used magneto-optical techniques to study the spin dynamics in semiconductor crystals and their nanostructures~\cite{awschalom2002,yakovlevCh6}. They exploit circularly polarized pump pulses to generate spin-oriented electrons and holes, which spin dynamics are detected by linearly polarized probe pulses via their Faraday or Kerr rotation. The techniques have been successfully used to investigate the coherent spin dynamics of electrons and holes in lead halide perovskite semiconductors and to measure $g$-factors, spin relaxation and spin dephasing times. Their power has been demonstrated for  bulk crystals of CsPbBr$_3$~\cite{belykh2019,Huynh2022PRB}, FA$_{0.9}$Cs$_{0.1}$PbI$_{2.8}$Br$_{0.2}$~\cite{kirstein2022}, MAPbI$_3$~\cite{kirstein2022Mapbi,huynh2022mapi}, FAPbBr$_3$~\cite{kirstein2024FAPbBr3}, MAPbBr$_3$~\cite{huynh2024}, and polycrystalline films of CsPbBr$_3$~\cite{grigoryev2021,jacoby2022}, MAPb(Cl,I)$_3$~\cite{odenthal2017}, MAPbI$_3$~\cite{garcia-arellano2021,garcia-arellano2022}, and FAPbI$_3$~\cite{lague2024} (FA = formamidinium). Universal dependences of the electron and hole $g$-factors on the band gap were found for these materials~\cite{kirstein2022uni,kopteva2023ex}. A collection of times characterizing the spin dynamics can be found in ref.~\onlinecite{kirstein2024FAPbBr3}. Commonly, the spin dephasing times are longer in single crystals, where the longest time for electrons of 11.5~ns was reported for FAPbBr$_3$~\cite{kirstein2024FAPbBr3} and for holes of 8~ns for MAPbBr$_3$~\cite{huynh2024}. In films, these times often are shorter than 1~ns, while recently comparably long times were reported for  MAPbI$_3$ films~\cite{garcia-arellano2021,garcia-arellano2022}: 4.4~ns for electrons and 7~ns for holes. For comparison, the longest times in MAPbI$_3$ single crystals are 11~ns for electrons and 6~ns for holes~\cite{huynh2022mapi}.

For the spin dynamics time scales approaching or exceeding the repetition period of the exciting pump laser pulses, spin accumulation can arise from the addition of spin dynamics induced by consecutive pulses. In order to gain comprehensive information on spin relaxation times in this case, not only measuring the spin dynamics, but also measuring the spin signals at a fixed pump-probe delay while scanning the magnetic field is instructive. Three such techniques exploiting the application of an external magnetic field have been developed and successfully used for conventional III-V and II-VI semiconductors:  resonant spin amplification (RSA)~\cite{kikkawa1998,Fokina2010,zhukov2012rsa}, spin mode-locking~\cite{greilich2006,yakovlevCh6}, and spin inertia~\cite{smirnov2018}. For lead halide perovskite semiconductors, the RSA method has not been reported so far. The spin mode-locking has the same origin of spin synchronization as the RSA effect, but appears in systems with a large dispersion of $g$-factors within an ensemble of carrier spins~\cite{yugova2012}. It was found for holes confined in CsPb(Br,Cl)$_3$ nanocrystals~\cite{Kirstein_SML_2023}. The spin inertia method was used for measuring the longitudinal spin relaxation times of carriers in CsPbBr$_3$~\cite{belykh2019}, MAPbI$_3$~\cite{kirstein2022Mapbi}, and FAPbBr$_3$~\cite{kirstein2024FAPbBr3}. 

In this paper we use these techniques to study the spin dynamics of charge carriers in MAPbI$_3$ single crystals at low temperatures in the orthorhombic phase. We find very long spin dephasing times reaching 21~ns for electrons, which allow us to enter the spin accumulation regime and use it for obtaining comprehensive information about the spin properties. Also the influence of the nuclear spins interacting with the carrier spins for the spin dynamics can be investigated by means of the polarization recovery (PR) effect and dynamic nuclear polarization (DNP).

\section{Results}
\label{Results}

\subsection{Optical properties}
\label{Optical}

\begin{figure*}[hbt]
 \includegraphics[width=\linewidth]{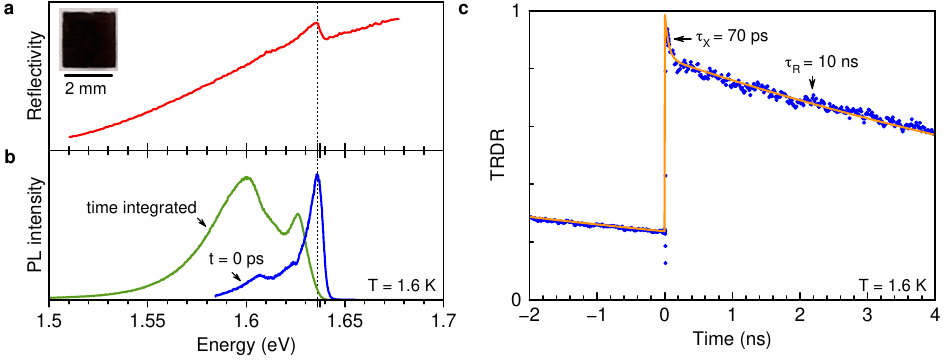}
  \caption{Optical properties of MAPbI$_3$ single crystal at $T=1.6$~K. 
	(a) Reflectivity spectrum showing a dispersively shaped resonance, with the exciton energy at 1.636\,eV marked by the dashed line. The inset gives a representative picture of the sample, adopted from ref.~\onlinecite{yang2022}. 
	(b) Photoluminescence measured for pulsed excitation right after the excitation pulse at $t=0$\,ps (blue) and time integrated (green).  
  (c) Time-resolved differential reflectivity (TRDR). Blue dots are the experimental data and orange line is a biexponential fit with the shorter decay time $\tau_\textnormal{X}=70$\,ps, provided by the exciton recombination, and the longer decay time $\tau_\textnormal{R}=10$\,ns, associated with recombination of spatially separated electrons and holes. The power densities of pump and probe are 1\,Wcm$^{-2}$. 
  }
\label{fig:intro}
\end{figure*}

The sample under study is a crystal of hybrid organic-inorganic perovskite MAPbI$_3$ with a thickness of 20~$\mu$m. In all the presented experiments, the light $\mathbf{k}$-vector is perpendicular to the sample surface, i.e., aligned along the z-axis, which coincides with the normal to the sample surface. The optical properties of the MAPbI$_{3}$ crystal at $T = 1.6$~K are presented in Figure~\ref{fig:intro}. In the reflectivity spectrum shown in Figure~\ref{fig:intro}a a pronounced exciton resonance is observed at $E_\text{X} = 1.636$~eV. The exciton binding energy in bulk MAPbI$_{3}$ is 16\,meV~\cite{galkowski2016}, and the estimated band gap energy is $E_\text{g} = 1.652$\,eV. 

Time-resolved and time-integrated photoluminescence (PL) spectra measured with a streak-camera are shown in Figure~\ref{fig:intro}b. Right after the laser excitation, the PL maximum is located at 1.636\,eV energy, corresponding to the emission of excitons, see blue spectrum. At this energy the signal shows a multi-exponential decay with two fast components characterized by times of 15\,ps and 85\,ps, corresponding to the exciton dynamics, and a much longer one with characteristic time of 520\,ps due to recombination of spatially-separated localized electrons and holes, see Ref.~\onlinecite{kopteva2025_OO_MAPI} for details. It is a common feature of bulk lead halide perovskite semiconductors that the emission lines from recombination of electron-hole pairs and of excitons are spectrally overlapping~\cite{kopteva2024OO,kudlacik2024,kirstein2022Mapbi,deQuilettes2019_si,herz2017}. Their contributions, however, can be separated by time-resolved techniques and also by polarized PL in magneto-optical experiments~\cite{kudlacik2024}. In the time-integrated PL spectrum (green line in Figure~\ref{fig:intro}b) the PL maximum is shifted towards lower energies down to 1.627\,eV. Also the low energy band with a maximum at 1.600~eV becomes most intense here.

We measure also the population dynamics of excitons and carriers by time-resolved differential reflectivity, see Figure~\ref{fig:intro}c. The dynamic evolution shows bi-exponential behavior. The short component with $\tau_{\rm X}=70$\,ps decay time can be associated with the exciton lifetime. The long-lived component has a larger amplitude and a decay time of $\tau_{\rm R} = 10$\,ns. As a result, it does not fully decay during the repetition period between the laser pulses of $T_\text{R}=13.2$\,ns so that a finite intensity is seen at negative delays. Note, that in this experiment we use linearly polarized light for the pump and probe beams and, therefore, measure the population dynamics which is insensitive to spin polarization. 

Note that recently we found in a similar, thin MAPbI$_3$ crystals optical spin orientation of excitons with a very high degree of 85\%, detected via polarized photoluminescence~\cite{kopteva2024_all_OO,kopteva2025_OO_MAPI}, which is robust against energy detuning of the excitation laser from the exciton resonance. It represents a common property of bulk lead halide perovskite semiconductors irrespective of their crystal symmetry (about cubic, tetragonal, orthorhombic)~\cite{kopteva2024_all_OO} and evidencing the absence of the Dyakonov-Perel spin relaxation mechanism.

\subsection{Spin dynamics of electrons and holes}
\label{Spin dynamics}

We use the time-resolved Kerr ellipticity (KE) technique to study the spin dynamics of electrons and hole. The spin polarization of electrons and holes is induced by circularly polarized pump pulses. According to the optical selection rules for lead halide perovskite semiconductors~\cite{kopteva2024OO}, a $\sigma^+$ polarized photon with angular momentum $+1$ generates a pair of an electron and a hole both having spin $+1/2$. Vice verse, a $\sigma^-$ polarized photon generates carriers with spin $-1/2$. The induced carrier spin polarization is detected via the Kerr rotation effect experienced by linearly polarized probe pulses, which are delayed in time relative to the pump pulses. For technical reasons, we measure in our experiments not directly the Kerr rotation, but the Kerr ellipticity, i.e., the presence of a small circularly polarized component in the reflected probe light, as it has maximum intensity at the exciton resonance~\cite{yugova2009}.  

\begin{figure*}[t!]
 \includegraphics[width=\linewidth]{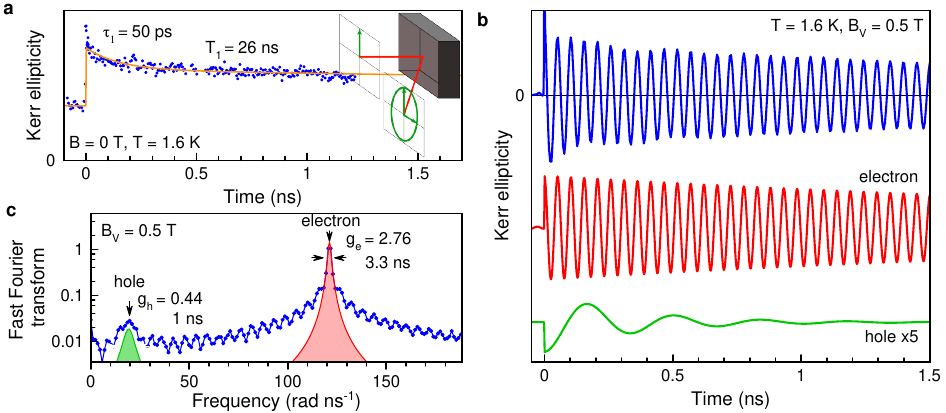}
  \caption{Spin dynamics in the MAPbI$_3$ single crystal measured at 1.637~eV. 
	(a) Spin dynamics at zero magnetic field (blue dots) and corresponding biexponential fit (orange line). The fit gives the spin lifetime $\tau_1 = 50$\,ps and the spin relaxation time $T_1 = 26$\,ns. Note that $T_1$ exceeds the laser pulse repetition rate $T_\textnormal{R}=13.2$\,ns, but due to the implemented experimental method of double modulation, the signal is nearly offset-free and the signal at negative time delays can be included in the fit. Sketch illustrating the Kerr ellipticity effect, converting the linearly polarized probe beam into an elliptical one. 
	(b) Spin dynamics in a magnetic field applied in the Voigt geometry: $B_{\rm V} = 0.5$\,T. $T=1.6$\,K, pump power of 1.6\,Wcm$^{-2}$. Below the vertically shifted electron and hole dynamics, decomposed by simulating the experimental dynamics, are shown. 
	(c) Fast Fourier transform (FFT) (blue line) of the signal shown in panel (b). The two Lorentzian fit curves give the electron and hole contributions. The denoted parameters are discussed in the text.}
\label{fig:dynamics}
\end{figure*}

The KE dynamics measured at zero magnetic field for 1.637\,eV photon energy, corresponding to the exciton resonance, are shown in Figure~\ref{fig:dynamics}a. The spin dynamics show two components, accordingly fitted with a biexponential decay functions. The short decay time of $\tau_1=50$\,ps is close to the exciton recombination time $\tau_{\rm X}$. Therefore, it can be assigned to the lifetime of the exciton spin $T_{\rm s,X}^{-1}=\tau_{\rm s,X}^{-1}+\tau_{\rm X}^{-1}$, which is limited by exciton recombination. Here $\tau_{\rm s,X}$ characterizes the exciton spin relaxation. Note that our studies of the exciton optical orientation effect and the magnetic-field-induced exciton circular polarization in similar MAPbI$_3$ crystals confirm that we are in a regime where $\tau_{\rm s,X} \gg \tau_{\rm X}$ holds at $T=1.6$~K for them~\cite{kopteva2025_OO_MAPI}.

The long-lived component in the KE dynamics of Figure~\ref{fig:dynamics}a has a decay time of $T_1=26$\,ns. Accordingly, we assign it to the longitudinal spin relaxation time of resident electrons and/or holes. Note that the $T_1$ time greatly exceeds the times involved in the exciton dynamics, confirming our interpretation of the long-living dynamics arising from spatially-separated resident carriers.

In a magnetic field applied in the Voigt geometry ($\textbf{B}_{\rm V} \perp \textbf{k}$), i.e., perpendicular to the induced spin polarization, the carrier polarization undergoes Larmor spin precession about the field direction~\cite{yakovlevCh6}. This results in oscillations in the KE dynamics, shown by the blue line in Figure~\ref{fig:dynamics}b for $B_{\rm V}=0.5$~T. The dynamics comprise two components, whose contributions we separate by a fit with Equation~\ref{eq:trke} and show in the same figure: the faster oscillating component with the Larmor precession frequency $\omega_{\text{L,e}} = 121.18$\,rad ns$^{-1}$ and the spin dephasing time $T_{2,\text{e}}^*=3.3$~ns belong to the electrons, while the slower oscillating one ($\omega_\text{L,h} = 19.23$\,rad ns$^{-1}$ and $T_{2,\text{h}}^*=0.35$~ns) arise from the holes. This identification of the components is done on basis of the known electron and hole $g$-factors for MAPbI$_3$ crystals, which are anisotropic and are in the ranges of $g_\text{e} = +2.46$ to $+2.98$ and $g_\text{h} = -0.28$ to $-0.71$~\cite{kirstein2022uni}.    

The Larmor precession frequency $\omega_\text{L}$ is proportional to the Zeeman splitting $E_\text{Z}$ which depends on the Land\'e $g$-factor according to:
\begin{equation}
E_\text{Z,e(h)} = \hbar \omega_\text{L,e(h)} = |g_\text{e(h)}| \mu_\textrm{B} B \,,
\label{eq:Zeeman}
\end{equation} %
with $\mu_\textrm{B}$ being the Bohr magneton. From the data presented in Figure~\ref{fig:dynamics}b, we evaluate $|g_\text{e}| = 2.76 \pm 0.10$ and $|g_\text{h}|=0.44 \pm 0.10$. Note that the $g$-factor signs cannot be identified from these experiments, but were determined for bulk MAPbI$_3$ using the dynamic nuclear polarization effect~\cite{kirstein2022}. 

Two frequencies, corresponding to the two oscillating components in the KE dynamics, are also well seen in the Fast Fourier Transform (FFT) spectrum (Figure~\ref{fig:dynamics}c). Here, the two peaks correspond to the hole and electron Larmor precession frequencies, and their full widths at half maximum (FWHM) give an estimate of their respective spin dephasing times: $\Delta \omega_\text{e} =1.91$\,rad\,ns$^{-1}$ ($T_{2,\text{e}}^*=2\pi/\Delta \omega_\text{e} = 3.3$\,ns) and $\Delta \omega_\text{h} = $ 6.03\,rad\,ns$^{-1}$ ($T_{2,\text{h}}^*=2\pi/\Delta \omega_\text{h}= 1$\,ns). 

Note the the amplitude of the hole spin contribution is much smaller than that of the electron, which might originate from a sample-specific drain of resident holes, see \cite{yang2022}. Commonly, comparable amplitudes for the electron and hole spin signals are observed in bulk lead halide perovskites~\cite{kirstein2022,belykh2019}.

\begin{figure*}[hbt]
 \includegraphics[width=\linewidth]{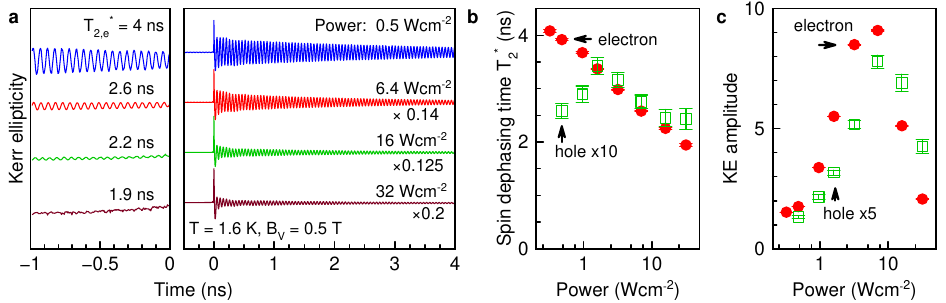}
  \caption{Spin dynamics at various pump powers, measured at $B_{\rm V}=0.5$~T. 
(a) KE dynamics for a set of different pump powers, but constant probe power of 0.3\,Wcm$^{-2}$. The sample is immersed in superfluid helium at $T = 1.6$~K. Most prominent at negative time delays (for the actual time delay, add $T_{\rm R} = 13.2$\,ns), the observed long-living spin oscillations vanish with increasing pump power. The signal amplitudes in the left panel are magnified by a factor of 40.
(b) Pump power dependence of the spin dephasing times $T_2^*$ of electrons and holes. The hole times are multiplied by a factor of 10 for better visibility.  
(c) Pump power dependence of the KE amplitudes. The hole amplitude is magnified by a factor of 5. The data in (b,c) are gathered via fits of the data set shown in (a) with Equation~\eqref{eq:trke}.
}
\label{fig:power}
\end{figure*}

In Figure~\ref{fig:power}a we show the KE dynamics measured at pump powers varied from 0.5\,Wcm$^{-2}$ to 32\,Wcm$^{-2}$. The dynamics' amplitudes are normalized. One sees that with increasing power the electron spin dephasing becomes faster, as seen in a pronounced manner from the decreasing oscillation amplitude at negative delays, which corresponds to a positive delay of about $T_\text{R}=13.2$\,ns. One can see in Figure~\ref{fig:power}b that $T_{2,\text{e}}^*$  shortens from 4\,ns to 2\,ns. The hole spin dephasing time is much shorter, being about 200\,ps, and weakly depends on pump power.

The KE amplitudes increase with the power growing up to about 6.4\,Wcm$^{-2}$ and then drop with a further power increase to 32\,Wcm$^{-2}$ (Figure~\ref{fig:power}c). We explain the decrease by electron heating and delocalization. In general, the carrier coherent spin dynamics are very sensitive to the bath temperature, as we will show for the studied sample below in Section~\ref{Temperature}. An increased carrier temperature is equivalent to an overall increased sample temperature. 

Most notably, in Figure~\ref{fig:power}a at negative time delays and for small pump power densities, the spin precession persists reflecting spin accumulation over several pump periods. We will exploit this accumulation effect in the next section.

\subsection{Resonant spin amplification}
\label{RSA}

Resonant spin amplification (RSA) has been found in n-type doped GaAs~\cite{kikkawa1998} and extended experimentally and theoretically to many conventional III-V and II-VI semiconductors and their quantum well heterostructures~\cite{awschalom2002,yakovlevCh6,yugova2012}. 

\begin{figure}[hbt]
 \includegraphics[width=1\linewidth]{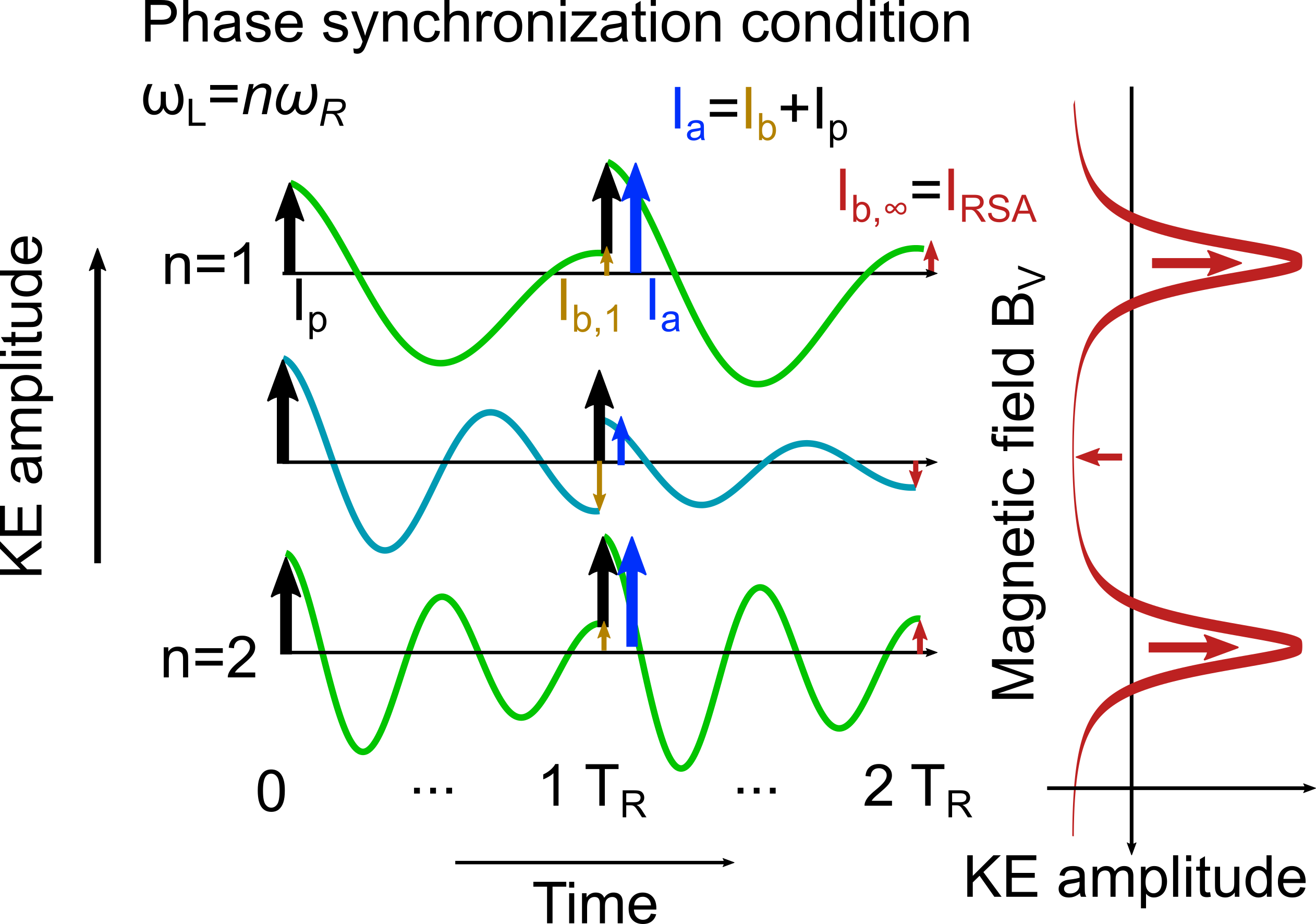}
  \caption{Resonant spin amplification effect. 
Sketch explaining how long-lived spin dynamics lead to spin accumulation and RSA. Three KE dynamics traces are shown with different Larmor precession frequencies. Black arrows indicate the periodic generation of spin polarization with each pump pulse arrival (${I}_{p}$). Green curves give precession with the Larmor frequency $\omega_{\rm L}$ being an integer number $n$ of the pulse repetition frequency $\omega_{\rm R}$, while the blue trace gives precession for $\omega_{\rm L} = 1.5 \omega_{\rm R}$. With arrival of the next pulse at $1 T_{\rm R}$, the remaining spin coherence (yellow arrows (${I}_{b,1}$)) adds constructively or destructively for the integer and half-integer relations cases, respectively, to the newly generated spin cohernce shown with the black arrows, resulting in the blue arrows (${I}_{a} = {I}_{b}+{I}_{p}$). The constructive case corresponds to the phase synchronization condition $\omega_{\rm L} = n \omega_{\rm R}$. This sequence of additive pump pulses repeats quasi-infinitely often leading to a strongly increased signal. The RSA signal (red curve) is obtained for continuously scanning the magnetic field, leading to a periodic match of the PSC condition. It is typically measured for small negative delays, i.e., shortly before the next pump pulse, marked by the red arrow in the dynamics. 
}
\label{fig:rsa0}
\end{figure}

In an RSA experiment, carrier spin coherence is excited by a pulsed laser with repetition period $T_{\rm R}$, which equals to $13.2$\,ns in our experiments. After the laser pulse, the photoinduced carrier spin polarization precesses about the magnetic field applied in the Voigt geometry, while decaying with the spin dephasing time $T_{2}^*$. In the case, when the spin dephasing time is larger than the repetition period ($T_{2}^*>T_{\rm R}$), the spin polarization does not fully decay up to the moment of the next pulse arrival. This next pulse can generate spin polarization which can interfere constructively with the decaying polarization. In this case the total polarization is amplified. If on the other hand the two polarizations are in anti-phase, destructive interference occurs minimizing the total polarization. Polarization accumulation occurs when the Larmor precession frequency is commensurate with the laser pulse repetition frequency, satisfying thereby the phase synchronization condition (PSC): 
\begin{equation} 
\omega_{\rm L} = n \omega_{\rm R} = n 2 \pi / T_{\rm R} \,.
 \label{eq:PRC} 
\end{equation}
Here, $n$ is an integer. A schematic presentation of the spin accumulation in the RSA process is shown in Figure~\ref{fig:rsa0}. 

A detailed theoretical consideration of the RSA effect can be found in ref.~\onlinecite{yugova2012}. Here, we present the key equations from this paper. Considering indistinguishable spins, after the first pump pulse action, the spin dynamics oscillate with the Larmor frequency, decaying with the spin relaxation time $\tau_\text{s}$. To describe the process of spin accumulation, the amplitude of the spin signal (in our case, this is Kerr ellipticity amplitude $A_{\rm KE}^{\rm RSA}$) is calculated as sum of the spin dynamics over an infinite number of laser pulses: 
\begin{eqnarray} \nonumber
A_{\rm KE}^{\rm RSA}(\omega_\text{L}(B), t) &=& \sum_{m=0}^{\infty} S_0 \exp[-(t + m T_{\rm R})/\tau_\text{s}] \times\\
 && \cos{[\omega_{\rm L}(t+ m T_{\rm R})]} \,.
 \label{eq:RSA1} 
\end{eqnarray}
Here $m$ is the number of the pulse and $S_0$ is the spin polarization induced by each pump pulse action. The summation over $m$ gives:
\begin{eqnarray} \nonumber
A_{\rm KE}^{\rm RSA}(\omega_\text{L}(B), t) =\frac{S_0}{2} \exp[-t/\tau_\text{s}] \times \\ 
 \frac{\exp[-T_{\rm R}/\tau_\text{s}]   \cos{(\omega_{\rm L} t)}-\cos{[\omega_{\rm L}(t+T_{\rm R})]}}{\cosh[T_{\rm R}/\tau_\text{s}]-\cos{(\omega_{\rm L} T_{\rm R})}} \,.
\label{eq:RSA} 
\end{eqnarray} 

\begin{figure*}[t!]
 \includegraphics[width=0.7\linewidth]{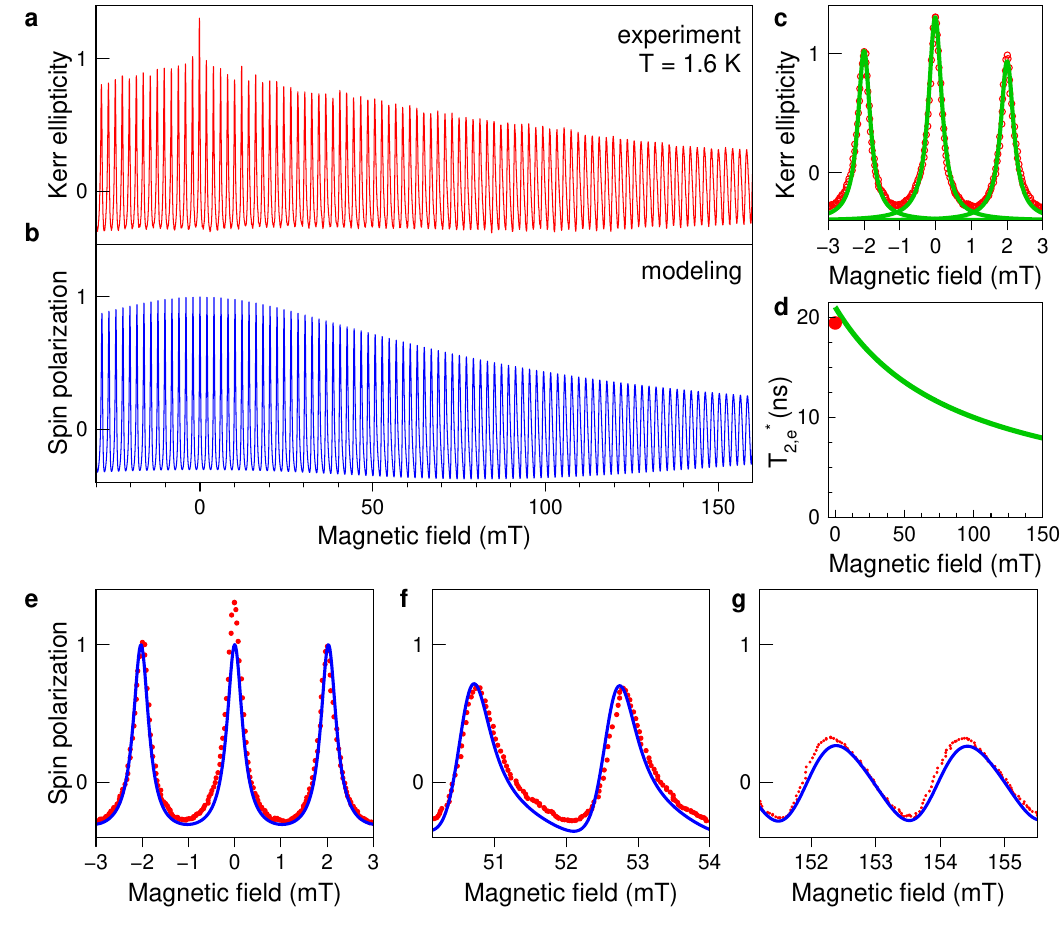}
  \caption{Resonant spin amplification of electrons in the MAPbI$_3$ crystal measured at 1.637~eV photon energy. 
	 (a) Magnetic field dependence of the Kerr ellipticity amplitude detected at the small negative delay of $t = -50$\,ps. The magnetic field is applied in Voigt geometry. The pump power is 0.5\,Wcm$^{-2}$ at $T=1.6$~K.   
	 (b) Simulated RSA signals using Equation~\eqref{eq:RSA} with the parameters $g_{\text{e},0} = 2.676$, $\Delta g_{\rm e} = 0.006$, and $\tau_\text{s} = T_{2,0}^* = 21$~ns.  
	 (c) The three RSA peaks around zero magnetic field (red dots give experimental data) and their fit by three Lorenzian functions (green lines) using Equation~\eqref{eq:RSA2}. 
		(d) Magnetic field dependence of the spin dephasing time $T_{2}^*$ (green line). We use Equation~\eqref{eq:spread} with the parameters $\Delta g_{\rm e} = 0.006$, and $T_{2,0}^* = 21$~ns. Red point is the value $T_{2,0}^* = 19.4$~ns obtained from the Lorentzian fit in panel (c).
		(e-g) Zoom into RSA peaks and their fit (blue line) with Equation~\eqref{eq:RSA}.
	}
\label{fig:rsa1}
\end{figure*}

Experimentally, for investigating the RSA effect it is favorable not to measure the time-resolved dynamics, but fix the time delay at a small negative value and scan the magnetic field while detecting the KE amplitude.  This allows one to perform measurements in weak magnetic fields, starting from zero field. Practically, the probe pulses are set in time to a value right before the arrival of the pump pulses. One describes this setting as a negative time delay of e.g. $t=-50$~ps, which in fact corresponds to a time delay close to the laser repetition period, i.e. $T_{\rm R}-50$~ps. The amplitude of the spin signal is measured as function of the external magnetic field. The RSA signal calculated with Equation~\eqref{eq:RSA} shows an oscillating behavior with sharp resonances separated in magnetic field by $\hbar \omega_\text{R}/\mu_\text{B} g$. The width of the resonances is determined by the spin relaxation time, see Figure~\ref{fig:rsa1}b. 

Equation~\eqref{eq:RSA} can be simplified to a Lorentzian form at zero delay time $t$ when the two conditions $|\omega_\text{L} T_\text{R} - 2\pi n|\ll1$ and $\tau_\text{s} \gg T_\text{R}$ are fulfilled:
\begin{eqnarray}
A_{\rm KE}^{\rm RSA}(\omega_{\rm L}(B)) \sim \frac{1}{(\omega_\text{L} T_\text{R} - 2\pi n)^2 + (T_\text{R}/\tau_\text{s})^2}.
\label{eq:RSA2} 
\end{eqnarray}
Equation~\eqref{eq:RSA2} can be used for the evaluation of the spin relaxation time when the dispersion of the $g$-factors is not significant and the RSA peaks remain sharp, which is fulfilled for the condition $\tau_\text{s} \gg T_\text{R}$. To include the $g$-factor dispersion in the model, Equation~\eqref{eq:RSA} should be averaged over the $g$-factor distribution, see ref.~\onlinecite{yugova2012} for details. However, this requires numerical calculations. An estimate of the $g$-factor dispersion can be obtained by fitting a sequence of RSA peaks by Lorentzians and interpret the width as dephasing time.

An experimentally measured RSA signal is shown in Figure~\ref{fig:rsa1}a. It exhibits periodic peaks with the period of $\hbar \omega_\text{R}/\mu_\text{B} g = 2$~mT. By fitting the peak around $B_\text{V} = 0$~T using Equation~\eqref{eq:RSA2}, we obtain $\tau_\text{s} = T_{2,0}^* = 19.4$~ns. The amplitude of the peaks decreases for higher magnetic fields due to the increasing significance of the $g$-factor dispersion ($\Delta g$), because the dispersion of Larmor frequencies ($\Delta \omega_\text{L}$) increases with magnetic field, following $\Delta g = \hbar \Delta \omega_\text{L}/\mu_\text{B} B_\text{V}$. To determine $\Delta g$, we calculate an RSA signal by integrating Equation~\eqref{eq:RSA} using a Gaussian distribution of $g$-factors. The magnetic field dependence of the RSA amplitude is shown in Figure~\ref{fig:rsa1}b. The model parameters used are $g_{\text{e},0} = 2.676$ (the median $g$-factor in the distribution), $\Delta g_{\rm e} = 0.006$, and $\tau_\text{s} = T_{2,0}^* = 21$~ns. The experimental and simulated curves show good agreement, which can be seen by comparing Figures~\ref{fig:rsa1}a,b, as well as the details of the RSA peaks shown in Figures~\ref{fig:rsa1}e-g. Note that the RSA peaks become broader and transform into a sinusoidal shape with increasing magnetic field due to the increasing contribution of the $g$-factor dispersion. The $g$-factor that we obtain from the modeling coincides with that of the electrons in MAPbI$_3$. Therefore, the long living spin coherence can be uniquely assigned to localized electrons. It is worth to note that the measured spin dephasing time of $T_{\rm 2,0,e}^*=21$~ns is the longest reported so far for lead halide perovskite semiconductors, see our comments in the introduction on available literature. We attribute this finding to the high structural quality of the studied thin MAPbI$_3$ single crystal, as well as to the favorite experimental conditions ($T=1.6$~K and low laser density). 

The spin dephasing time is decreasing with magnetic field due to the decreasing impact of the $g$-factor dispersion and can be calculated as~\cite{yakovlevCh6}: 
\begin{equation}
\frac{1}{T_2^*(B)} =  \frac{1}{T_{2,0}^*} + \frac{\Delta g \mu_{\rm B} B}{\hbar}.
\label{eq:spread}
\end{equation}
Using $\Delta g_{\rm e} = 0.006$ and $T_{2,0}^* = 21$~ns, we simulate this dependence and show it in Figure~\ref{fig:rsa1}d (green line). The red dot gives the $T_{2,0}^*=19.4$~ns value estimated from the Lorentzian fit of a single peak, which agrees with the value obtained from the RSA curve modeling. 

We emphasize that $\Delta g_{\rm e} = 0.006$ or $\Delta g_{\rm e}/g_{\rm e} = 0.2\%$ is a quite small value, evidencing the high homogeneity of the electron spin system. Let us compare it with the $g$-factor difference for electrons excited on the high and low energy flank of the 1.5\,ps pump pulse with finite spectral width. The electron $g$-factor depends on the band gap energy ($E_g$, which in our case can be considered as energy of the optical transition) as~\cite{kirstein2022uni} 
\begin{eqnarray}
g_{\rm e} &=& -\frac{2}{3} + \frac{4}{3}\frac{p^2}{m_0} \frac{1}{E_g} + \Delta g_{\rm e}^\prime \,.
 \label{eq:universal}
\end{eqnarray}
Here, $p$ is the interband matrix element of the momentum operator, $m_0$ the electron rest mass, $\Delta g_{\rm e}^\prime=-1$ the remote band contribution. In the lead halide perovskites, $\hbar p /m_0=6.8$~eV$\rm \AA$~\cite{kirstein2022uni}. For the energy range, covered by the pulse width of 1.5~meV we estimate a $g$-factor difference of about $0.004$, which is close to the $\Delta g_{\rm e}=0.006$ extracted from the RSA curve.  

\subsection{Polarization recovery}
\label{PRC}

In Faraday geometry, the external magnetic field is applied parallel to the pump light wavevector and, therefore, parallel to the generated carrier spin polarization. These carriers possess no average spin projection perpendicular to the field and do not undergo Larmor spin precession in the external magnetic field. Therefore, one expects that their spin dynamics are controlled by the longitudinal spin relaxation time $T_1$. 

\begin{figure*}[t!]
 \includegraphics[width=0.9\linewidth]{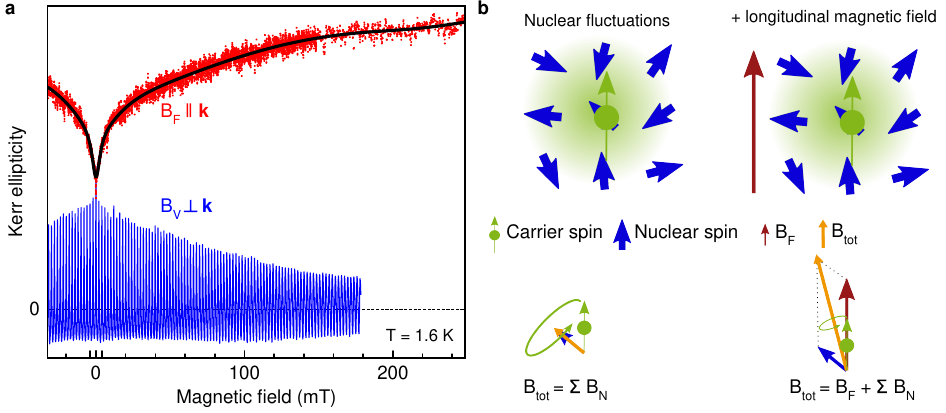}
  \caption{Polarization recovery. (a) Kerr ellipticity measured at the small negative time delay $t = -50$\,ps as function of the magnetic field scanned in Faraday geometry (red dots, PR curve) and in Voigt geometry (blue line, RSA). Black line is fit to the PR curve with a three-Lorentzian function of Equation~\eqref{eq:PR}. The excitation density of the pump is 0.5\,Wcm$^{-2}$ and of the probe is 0.35\,Wcm$^{-2}$.
 	(b) Left: Sketch of carrier spin (green), interacting with several randomly oriented nuclear spins (blue) within the carrier localization volume, causing spin relaxation (mainly by dephasing in the nuclear fluctuation field). Right: Applying a magnetic field (brown arrow) along the spin polarization, can overcome the nuclear fluctuations stabilizing the carrier spin polarization.}
\label{fig:PR}
\end{figure*}

However, at cryogenic temperatures, the nuclear spins play a significant role in the spin dynamics of localized carriers via the hyperfine interaction. It has been studied in detail for conventional III-V and II-VI semiconductors \cite{glazovbook} and also demonstrated for the lead halide perovskite crystals~\cite{kirstein2022,kirstein2022Mapbi,kudlacik2024}. Being localized, a carrier interacts with a finite number of nuclear spins, which have random spin orientations. The magnetic moment of the nuclear spin fluctuation acts on the carrier spin as an effective magnetic field, which is known as the nuclear Overhauser field. The nuclear fluctuations are also randomly oriented. Therefore, the carrier spin precesses around the perpendicular component of the nuclear Overhauser field, as sketched schematically in Figure~\ref{fig:PR}b.  At zero external field, this precession provides carrier spin relaxation, but this mechanism can be suppressed in an external magnetic field with strength exceeding that of the nuclear fields. Detailed considerations of this effect with accounting for the perovskite specifics can be found in ref.~\onlinecite{kudlacik2024}. Experimentally, it can be observed as polarization recovery (PR) effect, i.e., as an increases of the spin polarization signal in longitudinal magnetic field~\cite{glazovbook}. 

We measure the PR in the studied MAPbI$_3$ crystal by detecting the KE amplitude at a small negative delay of $-50$~ps. Its magnetic field dependence is shown by the red symbols in Figure~\ref{fig:PR}a, where also the RSA signal is shown for comparison. The KE amplitude increases by a factor of 2 with the magnetic field increasing to $B_{\rm F}=250$~mT. One can see that several processes are involved in this dependence. In order to quantify them, we fit the PR magnetic field dependence with a function composed of several Lorentzians~\cite{smirnov2020,kudlacik2024}:
\begin{equation}
A_{\rm KE}^{\rm PR} (B_{\rm F}) = A_{\rm KE, sat}-\sum_{i }{\frac{A_{i}}{ 1 + \left (\frac{B_{\rm F}}{\delta_{{\rm PR},i}}\right )^2}}.
\label{eq:PR}
\end{equation}
Here $A_{\rm KE, sat}$ is the saturation level of the KE amplitude at large magnetic fields and $\delta_{\textrm{PR},i}$ is the characteristic half width at half maximum (HWHM) of the Lorentzians. A fit of the PR experimental data with Equation~\eqref{eq:PR} is given by the yellow line in Figure~\ref{fig:PR}a. It shows that the PR signal consists indeed of three components with: $\delta_{\textrm{PR},1} = 3$\,mT, $\delta_{\textrm{PR},2} = 21$\,mT, and $\delta_{\textrm{PR},3} = 123$\,mT, and amplitudes $A_{\rm KE, sat}= 1$, $A_1 = 0.16$, $A_2 = 0.14$, and $A_3 = 0.28$.

By comparing these data with the recently published results for PR in FA$_{0.9}$Cs$_{0.1}$PbI$_{2.8}$Br$_{0.2}$, where we found for electrons $\delta_{\rm PR,e} = 5$~mT and for holes $\delta_{\textrm{PR,h}} = 19$~mT~\cite{kudlacik2024}, we assign the $\delta_\textrm{PR,1} = 3$~mT to localized electrons and the $\delta_\textrm{PR,2} = 21$~mT to localized holes in our sample. In the lead halide perovskites, the holes have a stronger hyperfine interaction with the nuclear spins compared to the electrons, which results in a broader PR curve. The origin of the process behind the broad signal with $\delta_{\textrm{PR},3} = 123$\,mT needs further clarification. We suggest that it is related to mechanisms which are not related to nuclear spins, but provide a $T_1(B_{\rm F})$ dependence for the carriers.     
   
We show in Section~\ref{Inertia} by spin inertia measurements that the longitudinal spin relaxation time $T_1$ in MAPbI$_3$ crystals is 20~ns at zero magnetic field and increases to 30~ns in $B_{\rm F}=20$~mT. Notably, this $T_1$ time is longer than the pump laser repetition period $T_{\rm R}=13.2$~ns in our experiment. In this case the photogenerated spin polarization does not fully relax between the laser pulses and spin accumulation can occur. Its effect can be described by the following equation: 
\begin{equation}
A_{\rm KE}(t) = \sum_{m=0}^{\infty}{S_0 \exp[-(t + m T_{\rm R})/T_1} ] \,.
\label{eq:sumPR}
\end{equation}
Further illustrations of the spin accumulation effect based on this modeling can be found in the Supporting Information.

\subsection{Spin inertia}
\label{Inertia}

To measure the longitudinal spin relaxation time $T_1$, the spin inertia technique can be employed~\cite{heisterkamp2015,zhukov2018si}. For that, the carrier spin polarization is driven by laser light, which can be pulsed or continuous-wave, with periodically circular polarization modulated between $\sigma^+$ and $\sigma^-$ with a modulation frequency $f_{\rm m}$. During each period, the spin polarization induced during the previous period is repolarized toward the opposite orientation. The repolarization occurs within the characteristic spin lifetime $T_{\rm s}$, which depends on the longitudinal spin relaxation time $T_1$, the carrier generation rate $G$, and the resident carrier concentration $n_0$:
\begin{equation}
\dfrac{1}{T_{\rm s}} = \frac{1}{T_1} +  \frac{G}{n_0} \, .
\label{eq:spininertia_time}
\end{equation} 
For a small generation rate, i.e., a weak excitation density, $T_{\rm s} \approx T_1$.

\begin{figure*}[t!]
 \includegraphics[width=\linewidth]{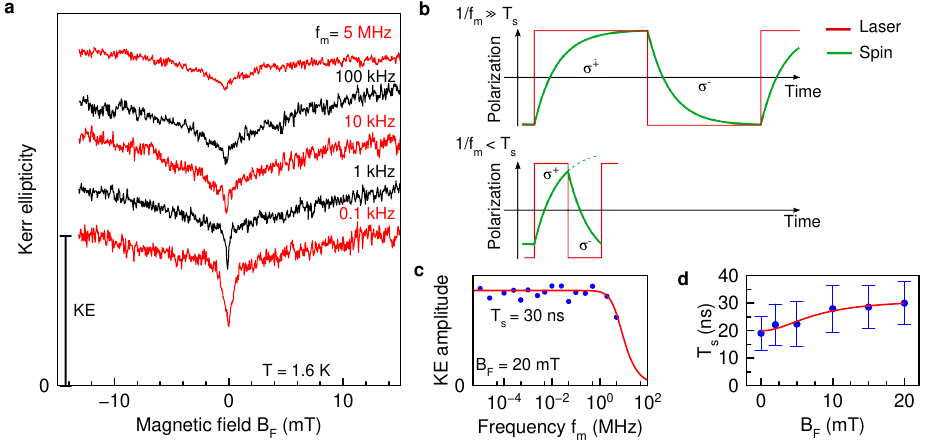}
  \caption{Spin inertia.
(a) Polarization recovery curves measured at the negative delay of $-50$\,ps as function of magnetic field scanned in Faraday geometry for different helicity modulation frequencies. All dependences, except the one for 0.1\,kHz, are shifted vertically for clarity. The excitation density of the pump is 0.5\,Wcm$^{-2}$ and that of the probe is 0.35\,Wcm$^{-2}$.
(b) Sketch of spin inertia. Upper diagram is for $1/f_{\rm m} \gg T_{\rm s}$ and lower one for $1/f_{\rm m}<T_{\rm s}$. The spin polarization is driven by a train of alternatively $\sigma^+$ and $\sigma^-$ polarized periods (red line). The spin polarization (green line) shows inertia, following the excitation polaization change with some delay characterized with the time $T_{\rm s}$. 
(c) KE amplitude in the saturation regime at $B_{\rm F} = 20$\,mT  as function of helicity modulation frequency $f_{\rm m}$  (blue symbols). Red line is a fit with Equation~\eqref{eq:spininertia} giving $T_{\rm s} = 30$\,ns.
(d) Magnetic field dependence of $T_{\rm s}$ (blue symbols) where the red line is a guide for the eye. 
}
\label{fig:spinin}
\end{figure*}

One can see in the diagrams of Figure~\ref{fig:spinin}b that if the polarization modulation period greatly exceeds $T_{\rm s}$ (i.e. $1/f_{\rm m}\gg T_{\rm s}$) the spin polarization reaches its maximal saturation value. However, for faster modulation ($1/f_{\rm m}<T_{\rm s}$) the maximal amplitude is not reached. Therefore, experimentally $T_{\rm s}$ can be evaluated by measuring the KE amplitude as function of $f_{\rm m}$ and fitting it with the form~\cite{heisterkamp2015}
\begin{equation}
A_{\rm KE}(f_\textrm{m}) \propto \frac{1}{\sqrt{1+(2\pi f_\textrm{m} T_{\rm s} )^2}} \, .
\label{eq:spininertia}
\end{equation}
The measurements can be performed at various magnetic fields applied in the Faraday geometry to obtain the $T_{\rm s}(B_{\rm F})$ dependence. 

For the studied MAPbI$_3$ sample, the measurements are performed at the small negative delay of $-50$~ps, where the electron spin polarization mainly contributes to the spin dynamics. Examples of PR curves measured for polarization modulation frequencies in the range from 0.1\,kHz up to 5\,MHz are shown in Figure~\ref{fig:spinin}a. Note that these PR curves are shifted vertically for clarity. The KE amplitude measured at $B_{\rm F} = 20$\,mT  as function of  $f_{\rm m}$ is shown in  Figure~\ref{fig:spinin}c. There is a clear decrease of the KE amplitude for the highest frequency of 5\,MHz. A fit with Equation~\eqref{eq:spininertia} yields $T_{\rm s} = 30$\,ns, which exceeds $T_{\rm R}=13.2$~ns, confirming the spin accumulation regime for the electrons. Note that we use a relatively low excitation density in this experiment, so that $T_{\rm s} \approx T_1$ and we can conclude that $T_1 \ge 30$~ns. 

The magnetic field dependence of $T_{\rm s}(B_{\rm F})$ in the field range up to 20~mT is given in Figure~\ref{fig:spinin}d. $T_{\rm s}(B_{\rm F})$ increases from 20\,ns at zero field up to 30\,ns at 20\,mT, in line with the PR dependence shown in Figure~\ref{fig:PR}a.

\subsection{Dynamic nuclear polarization}
\label{DNP}

The hyperfine interaction with the nuclear spins plays the dominant role in the spin relaxation of electrons and holes in lead halide perovskite semiconductors at cryogenic temperatures~\cite{kirstein2022,kudlacik2024,Kirstein_DNSS_2023}. The nuclear spins  have random orientations and provide efficient spin relaxation of the carriers~\cite{Meliakov_2024PRB}, while the spin relaxation of the nuclear spins takes a long time reaching seconds in perovskites. This can be used to preserve the spin orientation of carriers for a long time and, therefore, enhance spin accumulation effects. Also, adjustment of Larmor precession frequencies by the nuclear frequency focusing effect in the spin mode-locking regime~\cite{greilich2007} can be used to modify the coherent spin dynamics of carriers~\cite{Kirstein_SML_2023}.

\begin{figure*}[t!]
 \includegraphics[width=0.9\linewidth]{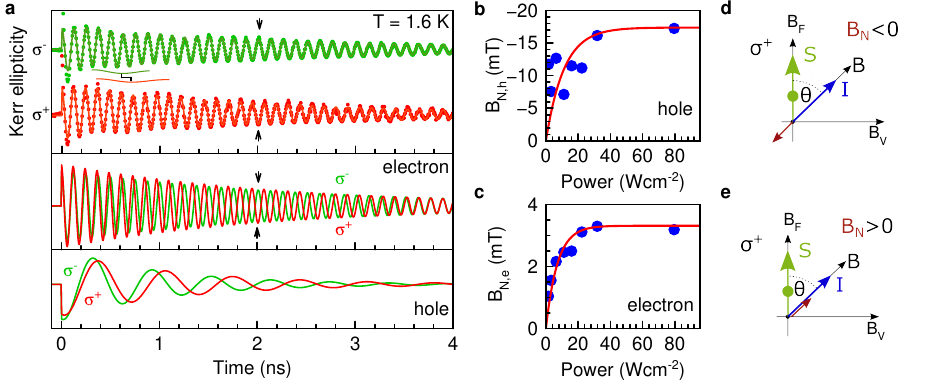}
  \caption{Dynamic nuclear polarization. (a) KE dynamics measured with $\sigma^+$ (red) or $\sigma^-$ (green) pump polarization at $B = \sqrt{(B_{\rm F}^2 + B_{\rm V}^2)} = 0.2$\,T, tilted by the angle $\theta = 60^\circ$ from the Faraday geometry. The pump power is 32\,Wcm$^{-2}$. The dynamics are fitted with Equation~\eqref{eq:trke} from which the electron and hole components are extracted and given below with corresponding colors. Note that the $\sigma^-$ data are multiplied by $-1$, to have the same signal sign at $t=0$ as the $\sigma^+$ data. DNP results in the emergence of the nuclear Overhauser field $B_{\rm N}$ acting on the carriers and shifting their Larmor precession frequency.	The considerable difference in the electron oscillation frequency is highlighted by the arrows at 2\,ns, showing opposite amplitudes. For the holes, it is evidenced as shift of an overtone-like modulation of the curves, as indicated at 0.5\,ns. 
	(b,c) Overhauser field $B_{\rm N}$ for the holes and the electrons, calculated according to Equation~\eqref{eq:DNP_2}. 
	(d,e) Sketches illustrating the DNP mechanism mediated by the holes and the electrons. Note that the difference in the Overhauser field sign for electrons and holes comes from the opposite signs of their $g$-factors. 
	}
\label{fig:DNP}
\end{figure*}

Spin polarized carriers can transfer their polarization to the nuclear spin system, inducing a dynamic nuclear polarization (DNP). In turn, the polarized nuclei act on the carrier spins via the nuclear Overhauser field, which can be considered as effective magnetic field inducing carrier Zeeman splitting. The nuclear Overhauser field in combination with the external magnetic field changes the Larmor precession frequency of the carriers and, therefore, can be detected in time-resolved KE experiments~\cite{kirstein2022}. This experiment requires a tilted geometry of the magnetic field, where the external field component perpendicular to the carrier spin polarization (i.e., perpendicular to the pump beam $k$-vector) is responsible for the Larmor precession, while the external field component parallel to the carrier spin polarization provides the spin transfer for achieving DNP. The induced DNP reads  
\begin{equation}
\braket{\mathbf{I}} = l \frac{4I(I+1)}{3} \frac{\mathbf{B}(\mathbf{B}\cdot\braket{\mathbf{S}_{\rm e(h)}})}{B^2} \,.
\label{eq:DNP}
\end{equation}
Here $\l$ is a leakage factor, $I$ is the nuclear spin, $\textbf{S}_{\rm e(h)}$ is the steady-state polarization of the carriers induced by optical orientation, for details see refs.~\onlinecite{kirstein2022,OptOr84}. 
 
The experimental demonstration of the DNP and its detection via time-resolved KE are shown in Figure~\ref{fig:DNP}a. Here the magnetic field of 0.2\,T is tilted by the angle of $\theta = 60^\circ$ from the Faraday geometry. The spin dynamics are measured for constant pump helicity, either $\sigma^+$ (red symbols and line) or $\sigma^-$ (green symbols and line). For better comparison, we invert the phase of the $\sigma^-$ pumped dynamics by multiplying them by $-1$. We fit the KE dynamics with Equation~\eqref{eq:trke} and plot the electron and hole components in the lower panels of Figure~\ref{fig:DNP}a. One can see, that the change of the pump helicity sign has an effect on the Larmor precession frequency both for electrons and for holes. This directly demonstrates the presence of a nuclear Overhauser field, which for the $\sigma^+$-polarized pump either adds to the external magnetic field, as in case of the electrons which Larmor frequency becomes higher, or subtract from it as in case of the holes, which frequency reduces. This difference arises from the opposite signs of the electron and hole $g$-factors in MAPbI$_3$. The diagrams in Figures~\ref{fig:DNP}d,e give more details of the DNP process involving the holes and the electrons.  

The nuclear Overhauser field can be evaluated from the difference in Larmor precession frequencies measured for $\sigma^+$ and $\sigma^-$ polarized pumps:
\begin{equation}
B_{\rm N,e(h)} = \dfrac{ \hbar(\omega_+ - \omega_-)}{2 g_{\rm e(h)} \mu_{\rm B}}.
\label{eq:DNP_2}
\end{equation}
where $\omega_\pm$ are the Larmor precession frequencies for a $\sigma^\pm$ pump. The evaluated Overhauser fields using $g_{\rm h}=-0.44$ and $g_{\rm e}=+2.76$ are plotted as function of the pump power in Figures~\ref{fig:DNP}b,c. Both for holes and electrons an initial increase converts to saturation. The saturation is due to the interplay between a higher carrier spin polarization with increasing pump power and an acceleration of the carrier spin relaxation, see Figure~\ref{fig:power}. Note that the Overhauser field is considerably stronger for holes than for electrons, which is common for the lead halide perovskites, where the spins of the lead ions mainly contribute to the hyperfine interaction~\cite{kirstein2022}.

\subsection{Impact of lattice temperature}
\label{Temperature}

\begin{figure*}[t!]
 \includegraphics[width=\linewidth]{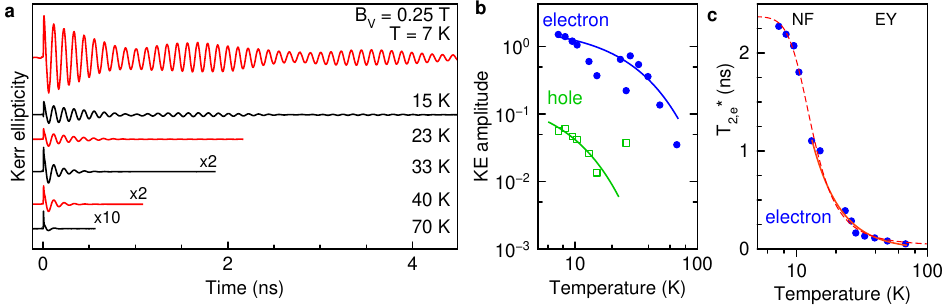}
  \caption{Temperature dependence of spin dynamics. (a) KE dynamics at different temperatures. The dynamics are fitted with Equation~\eqref{eq:trke} and the evaluated parameters are shown in panels (b,c).
(b) Temperature dependence of the amplitudes of the carrier spin components (dots). Lines are for the eye. Note the double logarithmic scale.
(c) Temperature dependence of the electron spin dephasing time (symbols). Fits are shown using Equation~\eqref{eq:EY} (solid line) for $T \geq 10$\,K and using an Arrhenius form according to Equation~\eqref{eq:arh} (dashed line).  
Excitation density of pump is 3.2\,Wcm$^{-2}$  and of probe 0.8\,Wcm$^{-2}$.}
\label{fig:temp}
\end{figure*}

The carrier spin dynamics depend sensitively on the lattice temperature. A set of KE dynamics in the temperature range from 7~K to 70~K is shown in Figure~\ref{fig:temp}a. With increasing lattice temperature the signal amplitude decreases rapidly and the spin dephasing accelerates. The dynamics are analyzed by fits with Equation~\eqref{eq:trke}. The resulting temperature dependences of the KE amplitude and the electron spin dephasing time ($T_{\rm 2,e}^*$) are presented in Figures~\ref{fig:temp}b,c. Note that above 23\,K the pump photon energy is tuned with temperature according to the law $E_\lambda/T = +0.5$\,meVK$^{-1}$ with an offset of 1.630\,eV, in order to accordingly achieve maximum signal.

The KE amplitude decreases strongly for growing temperature, see Figure~\ref{fig:temp}b. The spin dephasing time can be evaluated only for the electrons, as the accuracy is too low for the holes. A phenomenological Arrhenius-like activation function can describe well its temperature dependence:
\begin{equation}
\dfrac{1}{T_2^*(T)} = \dfrac{1}{T^*_{2,0}} + w \exp{\left(-\dfrac{E_A}{k_\textrm{B} T}\right)} \,.
\label{eq:arh}
\end{equation}
Here $E_A$ is the activation energy, $k_\textrm{B}$ is the Boltzmann constant, and $T^*_{2,0}$ is the temperature independent spin dephasing time. A fit shown by the dashed line gives $w = 0.032$\,nsK$^{-1}$, $T^*_{2,0} = 2.4$\,ns, and $E_A = 5$\,meV, in agreement with the lowest phonon modes in MAPbI$_3$~\cite{perez-osorio2018}. Qualitatively, the strong shortening of the $T_{\rm 2,e}^*$ time with increasing temperature can be explained by electron delocalization and the involvement of other spin relaxation mechanisms relevant for free carriers. 

As the Dyakonov-Perel spin relaxation mechanism is inefficient in bulk lead halide perovskites~\cite{kopteva2024_all_OO,kopteva2025_OO_MAPI}, the most relevant candidate is the Elliott-Yafet (EY) mechanism. It has the following temperature dependence~\cite{OptOr84}: 
\begin{equation}
\frac{1}{\tau_{s}^{EY}} = A_{\rm EY} \eta^2 \left ( \frac{1-\eta/2}{1-\eta/3}\right )^2 \left ( \frac{k_{\rm B} T}{E_g}\right )^2  \frac{1}{\tau_P} \, . 
\label{eq:EY}
\end{equation}
Here $\tau_s^{EY}$ is the spin relaxation time for the EY mechanism, $A_{\rm EY}$ is a constant, $\eta = \Delta_{\rm SO}/(E_g +\Delta_{\rm SO})$ with $\Delta_{\rm SO} \approx 1.5$\,eV being the spin-orbit coupling, $E_g$ is the band gap, and $1/\tau_P$ is the momentum scattering rate. Apparently, for a fit $A_{\rm EY}$ and $1/\tau_P$ are interdependent, thus we set $A_{\rm EY} = 1$ and get $\tau_P = 9.5$\,ns. The fit shown by the solid line in Figure~\ref{fig:temp}c is made for $T \geq 10$\,K, where the electrons are delocalized and can be treated as free.

For low temperatures the spin relaxation is dominated by the hyperfine interaction limiting the relaxation times. The general picture  of the contributing relaxation mechanisms is complex and depends strongly on the degree of localization of the carriers, which will be subject to further investigations.

\section{Discussion}
\label{Discussion}

To summarize the spin parameters obtained for MAPbI$_3$ thin crystals in this study, we collect them in Table~\ref{tab:sum}. The data for bulk MAPbI$_3$ crystals from ref.~\onlinecite{kirstein2022Mapbi} are given for comparison. The longitudinal spin relaxation times $T_1$ are comparable, lying in the range of about 30~ns. The hole spin dephasing times $T_{\rm 2,h}^*$ also do not differ strongly amounting to a few ns. However, the electron spin dephasing times $T_{\rm 2,e}^*$ differ strongly. $T_{\rm 2,e}^*$ is only 0.4~ns in bulk crystals and reaches a record value of 21~ns for the thin crystal. Note that the short electron spin dephasing time of 0.4~ns reported in ref.~\onlinecite{kirstein2022Mapbi} is not an inherent feature of MAPbI$_3$ bulk crystals, as recently  $T_{\rm 2,e}^*=11$~ns was reported for MAPbI$_3$ bulk crystals~\cite{huynh2022mapi}. The small dispersion of the carrier $g$-factors $\Delta g$ align with the high structural quality of bulk and thin crystals, confirmed by the narrow spectral lines of excitons in low-temperature reflectivity and PL. The maximum observed nuclear Overhauser field acting on the holes of $-17.2$\,mT in the thin crystals is larger but still close to the $-10.8$\,mT found in bulk crystals.

\begin{table}[hbt]
\caption{Spin dynamics parameters if electrons and holes in bulk and thin crystals of MAPbI$_3$ at $T=1.6$~K. the bulk values are taken from ref.~\onlinecite{kirstein2022Mapbi}.}
\begin{tabular}{c||c|c||c|c}
 & \multicolumn{2}{c||}{bulk} & \multicolumn{2}{c}{thin crystal} \\  
&hole&electron&hole & electron \\ \hline \hline
$A_{\rm h}/A_{\rm e}$ & \multicolumn{2}{c||}{$\approx$ 1/1} & \multicolumn{2}{c}{$\approx$ 1/5} \\
$T_1$ (ns) & $\leq 37$ &$\leq 37$ & $-$ & 30 \\
$T_{2,max}^*$ (ns) & 2.7 & 0.4 & $0.8$ & 21 \\
$\Delta g$ & $0.003-0.008$ & $0.03-0.04$& $-$ & 0.006 \\
DNP$_{\rm max}$ (mT) & $-10.8$&$+1.7$& $-17.2$ & $+3.3$ \\
\end{tabular}
\label{tab:sum}
\end{table} 

Interestingly, the hole spin signal is much weaker in thin crystals as compared to bulk, which might be related to a reduced hole abundance in thin single crystals~\cite{yang2022}. This in turn is an important piece of information to unravel the origin of the simultaneous hole and electron signals in perovskite crystals. 

\section{Conclusion}
\label{Conclusion}

To summarize, investigation of the long-lived carrier spin dynamics in thin MAPbI$_3$ single crystals allows us to demonstrate spin accumulation effects with conventional pulsed lasers operating at 76\,MHz repetition rate. The resonant spin amplification, spin inertia, and polarization recovery techniques, as established methods for studying the spin physics in conventional III-V and II-VI semiconductors, are suitable tools also for the lead halide perovskite semiconductors. A long longitudinal spin relaxation of 30~ns and a electron spin dephasing time of 21~ns are measured in the MAPbI$_3$ thin crystals. We are convinced that similarly long times can be reached in high quality lead halide perovskite crystals with other chemical compositions. We propose that the applied magneto-optical techniques based on spin accumulation will significantly advance the understanding of the spin dynamics in perovskite semiconductors, particularly in experiments involving laser systems with GHz repetition rates. All these results position the lead halide perovskites as promising functional materials for spintronics and information technologies, which potential can be greatly extended moving towards perovskite nanocrystals and two-dimensional materials.

\section{Experimental Section}

\textit{Samples:} The investigated MAPbI$_3$ thin single crystals were synthesized utilizing PI$_2$ and MAI perovskite precursors. These precursors were introduced between two polytetrafluoroethylene-coated glass substrates and subjected to a gradual thermal treatment, reaching a temperature of 120$^\circ$C. A precursor molar ratio of PI$_2$:MAI of 4:7 was employed. The thin single crystals exhibit a square morphology within the (001) crystallographic plane, with a thickness of approximately 20$\,\mu$m. At ambient temperature, the MAPbI$_3$ thin crystals have a tetragonal crystal structure, characterized by an out-of-plane tetragonal [001] axis. At cryogenic temperatures below 160~K the crystal structure changes to orthorhombic. The sample code is M2-5.	The sample has a size of $2 \times 2 \times 0.02$\,mm. An X-ray diffraction (XRD) analysis performed at room temperature confirmed the crystallographic orientation of the sample surface normal which is collinear to the light wave vector, $\textbf{k} \parallel  [001]$ (c-axis). 

\textit{Magneto-Optical Measurements:} For magneto-optical experiments the sample is placed in a cryostat allowing a temperature variable from 1.6\,K up to 300\,K. For $T=1.6$\,K the sample is immersed in superfluid helium, while for temperatures in the range from 4.2\,K to 300\,K the sample is held in helium cooling gas. The cryostat is equipped with a vector magnet composed of three superconducting split coils oriented orthogonal to each other. This allows us to apply magnetic fields up to 3\,T in any direction. We use a Cartesian coordinate system with the z-axis collinear to the sample surface normal (i.e. to the c-axis) and to the light wave vector $\textbf{k}$. The magnetic field collinear to the z-axis corresponds to the Faraday geometry ($\textbf{B}_{\rm F} \parallel \textbf{k}$). The magnetic field oriented in the plane perpendicular to $\textbf{k}$ corresponds to the Voigt geometry ($\textbf{B}_{\rm V} \perp \textbf{k}$). Note that the 3D vector magnet allows also for a precise compensation of the residual fields. 

\textit{Photoluminescence and Reflectivity:} Time-resolved photoluminescence (PL) is excited by a 200 fs-pulsed Ti:Sa laser emitting at the photon energy of 1.771\,eV (700\,nm). For detection we use a streak camera attached to an 0.5-m spectrometer, providing a time-resolution of about 20~ps. For the reflectivity measurements a halogen lamp is used and the reflected light is coupled via a fiber into an 0.5\,m spectrometer, equipped with a Peltier-cooled silicon charge-coupled device (CCD). 

\textit{Time-Resolved Kerr Ellipticity:} The coherent spin dynamics of electrons and holes interacting with the nuclear spins are measured by a degenerate pump-probe technique, where the pump and the probe have the same photon energy~\cite{yakovlevCh6}. A titanium-sapphire laser generates 1.5\,ps long pulses with a spectral width of about $1$\,nm (about 1.5\,meV) at a pulse repetition rate of 76\,MHz (repetition period $T_\text{R}=13.2$\,ns). The output photon energy is tuned to be about at the exciton resonance in the reflectivity spectrum to achieve maximum Kerr ellipticity signal at 1.637\,eV (757.3\,nm). The laser output is split into the pump and probe beams. The probe pulses are delayed relative to the pump pulses by a double-pass mechanical delay line with one meter length. The pump and probe beams are modulated using a photo-elastic modulator (PEM) for the probe and an electro-optical modulator (EOM) for the pump. The probe beam is always linearly polarized with the amplitude modulated at the frequency of 84\,kHz, with the PEM set to the $\lambda/2$ mode and combined with a Glan prism. The pump and probe beams are focussed on the sample, with pump spot diameter being 200\,$\mu$m and the probe spot size being slightly smaller.

The pump beam is either helicity modulated between $\sigma^+$ and $\sigma^-$ circular polarizations, or amplitude modulated with fixed helicity, either $\sigma^+$ or $\sigma^-$, in the frequency range from 0 to 5\,MHz. In all cases $f_{\rm m}$ refers to the helicity modulation frequency. Amplitude modulation can in effect be considered as 0~Hz helicity modulation, as the signal is independent of the bare amplitude modulation frequency. In the experiment, typically 20~Hz to 100~kHz helicity and amplitude modulation frequencies are used. The polarization of the reflected probe beam is analyzed with respect to the rotation of its elliptical polarization (Kerr ellipticity) using a balanced photodiode with a lock-in technique. The time-resolved Kerr ellipticity dynamics are analyzed via fits with the equation 
\begin{equation}
A_{\rm KE} (t) = \sum_{i=e,h}{A_i \cos{(\omega_{{\rm L},i} t)} \exp{(-t/T_{2,i}^*)}} .
\label{eq:trke}
\end{equation}
Here $i=e,h$ labels the electron and hole components, $A_i$ is the component specific amplitude, $\omega_{{\rm L},i}$ is the Larmor precession frequency, $T_{2,i}^*$ is the spin dephasing time of the carrier spin ensemble. For simplicity the $i$ index is not used, if not explicitly needed. Note, that $T_2^* \leq T_2$, where $T_2$ is the spin coherence time of the individual carriers.  \\


\textbf{Supporting Information.}
Additional information on the modeling of spin accumulation effects.

\textbf{Conflict of Interest.}
The authors declare no conflict of interest.

\textbf{Data Availability Statement.}
The data that support the findings of this study are available from the corresponding author upon reasonable request.

\textbf{Keywords.}
Lead halide perovskite crystals, carrier spin dynamics, time-resolved Kerr ellipticity, resonant spin amplification, spintronics.

\textbf{AUTHOR INFORMATION}

{\bf Corresponding Authors} \\
Erik Kirstein, Email: erik.kirstein@tu-dortmund.de\\
Dmitri R. Yakovlev,  Email: dmitri.yakovlev@tu-dortmund.de\\

\textbf{ORCID}\\
Erik~Kirstein:        0000-0002-2549-2115 \\
Dmitri R. Yakovlev:   0000-0001-7349-2745\\
Evgeny~A.~Zhukov:     0000-0003-0695-0093   \\
Nataliia E. Kopteva:   0000-0003-0865-0393 \\
Bekir Turedi:       0000-0003-2208-0737\\
Maksym V. Kovalenko:  0000-0002-6396-8938\\
Manfred Bayer:        0000-0002-0893-5949 \\

{\bf Notes}\\
The authors declare no competing financial interests.

\section{Acknowledgments}

The authors are thankful to M. M. Glazov, K. V. Kavokin, D. S. Smirnov, and M. Kotur for fruitful discussions. We acknowledge the financial support by the Deutsche Forschungsgemeinschaft via the SPP2196 Priority Program (project YA 65/28-1, no. 527080192). N.E.K. acknowledges support of the Deutsche Forschungsgemeinschaft (project KO 7298/1-1, no. 552699366). The work at ETH Z\"urich (B.T. and M.V.K.) was financially supported by the Swiss National Science Foundation (grant agreement 200020E 217589, funded through the DFG-SNSF bilateral program) and by ETH Z\"urich through ETH+ Project SynMatLab.

\renewcommand{\i}{\ifr}
\let\oldaddcontentsline\addcontentsline
\renewcommand{\addcontentsline}[3]{}




\begin{thebibliography}{99}

\bibitem{vardeny2022} Z.~V.~Vardeny, and M.~C.~Beard (eds.), \textit{Hybrid Organic Inorganic Perovskites: Physical Properties and Applications}, World Scientific, \textbf{2022}.

\bibitem{vinattieri2021} A.~Vinattieri, and G.~Giorgi (eds.), \textit{Halide Perovskites for Photonics}, AIP Publishing, Melville, New York, \textbf{2021}.

\bibitem{Martinez2023_book} J. P. Martinez-Pastor, P. P. Boix, and G. Xing (eds.), \textit{Halide Perovskites for Generation, Manipulation and Detection of Light}, Elsevier, \textbf{2023}.

\bibitem{privitera2021} A. Privitera, M. Righetto, F. Cacialli, and M. K. Riede,
Perspectives of organic and perovskite-based spintronics,
\textit{Adv. Optical Mater.}  \textbf{2021}, 2100215.

\bibitem{wang2019} 
J. Wang, C. Zhang, H. Liu, R. McLaughlin, Y. Zhai, S. R. Vardeny, X. Liu, S. McGill, D. Semenov, H. Guo, R. Tsuchikawa, V. V. Deshpande, D. Sun,  and Z. V. Vardeny, 
Spin-optoelectronic devices based on hybrid organic-inorganic trihalide perovskites.
\textit{Nat. Commun.} \textbf{2019}, \textit{10}, {129}.

\bibitem{kim2021} 
Y.-H. Kim, Y. Zhai, H. Lu, X. Pan, C. Xiao, E. A. Gaulding, S. P. Harvey, J. J. Berry, Z.~V. Vardeny, J. M. Luther,  and M. C. Beard, 
Chiral-induced spin selectivity enables a room-temperature spin light-emitting diode.
\textit{Science} \textbf{2021},	\textit{371}, {1129}.

\bibitem{kirstein2022uni}   
E.~Kirstein, D.~R.~Yakovlev, M.~M.~Glazov, E.~A.~Zhukov, D.~Kudlacik, I.~V.~Kalitukha, V.~F.~Sapega, G.~S.~Dimitriev, M.~A.~Semina, M.~O.~Nestoklon, E.~L.~Ivchenko, N.~E.~Kopteva, D.~N.~Dirin, O.~Nazarenko, M.~V.~Kovalenko, A.~Baumann, J.~H\"ocker, V.~Dyakonov, and M.~Bayer,
The Land\'e factors of electrons and holes in lead halide perovskites: universal dependence on the band gap.
\textit{Nat. Commun.} \textbf{2022}, \textit{13}, {3062}.

\bibitem{kopteva2024OO}  N. E. Kopteva, D. R. Yakovlev, E. Yalcin, I. A. Akimov, M. O. Nestoklon, M. M. Glazov, M. Kotur, D. Kudlacik, E. A. Zhukov, E. Kirstein, O. Hordiichuk, D. N. Dirin, M. V. Kovalenko, and M. Bayer,
Highly-polarized emission provided by giant optical orientation of exciton spins in lead halide perovskite crystals, \textit{Adv. Sci.} \textbf{2024}, \textit{11}, 2403691.

\bibitem{awschalom2002} D.~D.~Awschalom, D.~Loss, and N.~Samarth (eds.), \emph{Semiconductor Spintronics and Quantum Computation} (Springer, Berlin, \textbf{2002}).

\bibitem{yakovlevCh6} D.~R.~Yakovlev and M.~Bayer, in: \emph{Spin Physics in Semiconductors}, 2nd edition, edited by M. I. Dyakonov (Springer International Publishing, \textbf{2017}), Chap. 6, p. 155.

\bibitem{belykh2019} V. V. Belykh, D. R. Yakovlev, M. M. Glazov, P. S. Grigoryev, M. Hussain, J. Rautert, D. N. Dirin, M. V. Kovalenko, and M. Bayer, 
Coherent spin dynamics of electrons and holes in CsPbBr$_3$ perovskite crystals.
\textit{Nat. Commun.} \textbf{2019}, \textit{10}, {673}.
	
\bibitem{Huynh2022PRB} U. N. Huynh, T. Feng, D. R. Khanal, H. Liu, P. Bailey, R. Bodin, P. C. Sercel, J. Huang, and Z. V. Vardeny,
Transient and steady state magneto-optical studies of the CsPbBr$_3$ crystal,  
\textit{Phys. Rev. B} \textbf{2022}, \textit{106}, 094306. 

\bibitem{kirstein2022}  
E. Kirstein, D. R. Yakovlev, M. M. Glazov, E. Evers, E. A. Zhukov, V. V. Belykh, N. E. Kopteva, D. Kudlacik, O. Nazarenko, D. N. Dirin, M. V. Kovalenko, and M. Bayer,
Lead-dominated hyperfine interaction impacting the carrier spin dynamics in halide perovskites.
\textit{Adv. Mater.} \textbf{2022}, \textit{34}, 2105263.

\bibitem{kirstein2022Mapbi} E. Kirstein, D. R. Yakovlev, E. A. Zhukov, J. H\"ocker, V. Dyakonov, and M. Bayer,
Spin dynamics of electrons and holes interacting with nuclei in MAPbI$_3$ perovskite single crystals.
\emph{ACS Photonics} \textbf{2022}, \textit{9}, 1375--1384.

\bibitem{huynh2022mapi} 
U. N. Huynh, Y. Liu, A. Chanana, D. R. Khanal, P. C. Sercel, J. Huang, and Z. V. Vardeny,
Transient quantum beatings of trions in hybrid organic tri-iodine perovskite single crystal.  
\textit{Nat. Commun.} \textbf{2022}, \textit{13}, 1428. 

\bibitem{kirstein2024FAPbBr3} 
E. Kirstein, E. A. Zhukov, D. R. Yakovlev, N. E. Kopteva, E. Yalcin, I. A. Akimov, O. Hordiichuk, D. N. Dirin, M. V. Kovalenko, and M. Bayer. 
Coherent carrier spin dynamics in FAPbBr$_3$ perovskite crystals.  
\textit{J. Phys. Chem. Lett.} \textbf{2024}, \textit{15}, 2893-2903.

\bibitem{huynh2024} 
U. N. Huynh, R. Bodin, X. Pan, P. Bailey, H. Liu, S. McGill, D. Semenov, P. C. Sercel, and Z. V. Vardeny,
Hybrid organic/inorganic perovskite: The case of methylammonium lead bromide,  
\textit{Phys. Rev. B} \textbf{2024}, \textit{109}, 014316.
	
\bibitem{grigoryev2021} P. S. Grigoryev, V. V. Belykh, D. R. Yakovlev, E. Lhuillier, and M. Bayer,
Coherent spin dynamics of electrons and holes in CsPbBr$_3$ colloidal nanocrystals. 
\textit{Nano Lett.} \textbf{2021}, \textit{21}, 8481-8487.
	
\bibitem{jacoby2022}  L. M. Jacoby, M. J. Crane, and D. R. Gamelin, 
Coherent spin dynamics in vapor-deposited CsPbBr$_3$ perovskite thin films,
\textit{Chem. Mater.} \textbf{2022}, \textit{34}, 1937-1945. 

\bibitem{odenthal2017} 
P. Odenthal, W. Talmadge, N. Gundlach, R. Wang, C. Zhang, D. Sun, Z.-G. Yu, V. Z. Vardeny, and Y. S. Li,  
Spin-polarized exciton quantum beating in hybrid organic--inorganic perovskites.
\textit{Nature Physics} \textbf{2017}, \textit{13}, \textit{894}.
	
\bibitem{garcia-arellano2021}
G. Garcia-Arellano, G. Tripp\'e-Allard, L. Legrand, T. Barisien, D. Garrot, E. Deleporte, F. Bernardot, C. Testelin, and M. Chamarro,  
Energy tuning of electronic spin coherent evolution in methylammonium lead iodide perovskites. 
\textit{J. Phys. Chem. Lett.} \textbf{2021}, \textit{12}, \textrm{8272}.

\bibitem{garcia-arellano2022}
G. Garcia-Arellano, G. Tripp\'e-Allard, T. Campos, F. Bernardot, L. Legrand, D. Garrot, E. Deleporte, C. Testelin, and M.  Chamarro, 
Unexpected anisotropy of the electron and hole Land\'e g-factors in perovskite CH$_3$NH$_3$PbI$_3$ polycrystalline films. \textit{Nanomaterials} \textbf{2022}, \textit{12}, 1399. 

\bibitem{lague2024}  G. Lag\"ue, F. Bernardot, V. Guilloux, L. Legrand, T. Barisien, J. S\'anchez-Diaz, S. Galve-Lahoz, I. Saidi, K. Boujdaria, J. P. Martinez-Pastor, C. Testelin, I. Mora-Ser\'o, and M. Chamarro, 
Spin coherence and relaxation dynamics of localized electrons and holes in FAPbI$_3$ films,
\textit{ACS Photonics} \textbf{2024}, \textit{11}, 2770-2775. 

\bibitem{kopteva2023ex}
N. E. Kopteva, D. R. Yakovlev, E. Kirstein, E. A. Zhukov, D. Kudlacik, I. V. Kalitukha, V. F. Sapega, D. N. Dirin, M. V. Kovalenko, A. Baumann, J. H\"ocker, V. Dyakonov, S. A. Crooker, and M. Bayer, 
Weak dispersion of exciton Land\'e  factor with band gap energy in lead halide perovskites: Approximate compensation of the electron and hole dependences, 
\textit{Small} \textbf{2024}, \textit{16}, 2300935.

\bibitem{kikkawa1998} J. M. Kikkawa and D. D. Awschalom,
Resonant spin amplification in n-type GaAs,
\textit{Phys. Rev. Lett.} \textbf{1998}, \textit{80}, 4313.

\bibitem{Fokina2010} 
L. V. Fokina, I. A. Yugova, D. R. Yakovlev, M. M. Glazov, I. A. Akimov, A. Greilich, D. Reuter, A. D. Wieck, and M. Bayer, 
Spin dynamics of electrons and holes in InGaAs/GaAs quantum wells at millikelvin temperatures. 
\emph{Phys. Rev. B} \textbf{2010}, \textit{81}, 195304. 

\bibitem{zhukov2012rsa} 
E. A. Zhukov, O. A. Yugov, I. A. Yugova, D. R. Yakovlev, G. Karczewski, T. Wojtowicz, J. Kossut, and M. Bayer,
Resonant spin amplification of resident electrons in CdTe/(Cd,Mg)Te quantum wells subject to tilted magnetic fields. 
\emph{Phys. Rev. B} \textbf{2012}, \textit{86}, 245314. 

\bibitem{greilich2006} A. Greilich, D. R. Yakovlev, A. Shabaev, Al. L. Efros, I. A. Yugova, R. Oulton, V. Stavarache, D. Reuter, A. Wieck, and M. Bayer,
Mode locking of electron spin coherences in singly charged quantum dots.
\textit{Science} \textbf{2006}, \textit{313}, 341.

\bibitem{smirnov2018} D. S. Smirnov, E. A. Zhukov, D. R. Yakovlev, E. Kirstein, M. Bayer, and A. Greilich,
Theory of spin inertia in singly charged quantum dots,
\textit{Phys. Rev. B} \textbf{2018}, \textit{98}, 125306.        

\bibitem{Kirstein_SML_2023} E. Kirstein, N.~E. Kopteva, D.~R. Yakovlev, E.~A. Zhukov, E.~V. Kolobkova, M.~S. Kuznetsova, V.~V. Belykh, I.~A. Yugova, M.~M. Glazov, M.~Bayer, and A. Greilich.
Mode locking of hole spin coherences in CsPb(Cl,Br)$_3$ perovskite nanocrystals.  
\emph{Nat. Commun.} \textbf{2023}, \textit{14}, 699.

\bibitem{yugova2012} I. A. Yugova, M. M. Glazov, D. R. Yakovlev, A. A. Sokolova, and M. Bayer,
Coherent spin dynamics of electrons and holes in semiconductor quantum wells and quantum dots under periodical optical excitation: Resonant spin amplification versus spin mode locking.
\textit{Phys. Rev. B} \textbf{2012}, \textit{85}, 125304.  

\bibitem{kopteva2024_all_OO}  N. E. Kopteva, D. R. Yakovlev, E. Yalcin, I. A. Akimov, M. O. Nestoklon, M. M. Glazov, M. Kotur, D. Kudlacik, E. A. Zhukov, E. Kirstein, O. Hordiichuk, D. N. Dirin, M. V. Kovalenko, and M. Bayer, 
Effect of crystal symmetry of lead halide perovskites on the optical orientation of excitons,
\textit{Advanced Science}  \textbf{2025}, in press DOI: 10.1002/advs.202416786.
\textit{CondMat ArXive} 16 November \textbf{2024}, https://doi.org/10.48550/arXiv.2411.03764. 

\bibitem{kopteva2025_OO_MAPI}  N. E. Kopteva, D. R. Yakovlev, E. Yalcin, I. A. Akimov, M. Kotur, B. Turedi, D. N. Dirin, M. V. Kovalenko, and M. Bayer, 
 Optical orientation of excitons and carriers in perovskite MAPbI$_3$ single crystal in orthorhombic phase, 
\textit{CondMat ArXive} 07 February \textbf{2025}.  http://arxiv.org/abs/2502.04902.
	
\bibitem{galkowski2016} K. Galkowski, A. Mitioglu, A. Miyata, P. Plochocka, O. Portugall, G. E. Eperon, J. T.-W. Wang, T. Stergiopoulos, S. D. Stranks, H. J. Snaith,  and R. J. Nicholas, 
Determination of the exciton binding energy and effective masses for methylammonium and formamidinium lead tri-halide perovskite semiconductors.
\textit{Energy Environ. Sci.} \textbf{2016}, \textit{9}, {962-970}.

\bibitem{yang2022} 
C. Yang, J. Yin, H. Li, K. Almasabi, L. Guti\'errez-Arzaluz, I. Gereige, J.-L. Br\'edas, O. M. Bakr, and O. F. Mohammed, 
Engineering surface orientations for efficient and stable hybrid perovskite single-crystal solar cells. 
\emph{ACS Energy Lett.} \textbf{2022}, \textit{7}, 1544–1552. 

\bibitem{kudlacik2024} 
D.~Kudlacik, N.~E.~Kopteva, M.~Kotur, D.~R.~Yakovlev, K.~V.~Kavokin, C.~Harkort, M.~Karzel, E.~A.~Zhukov, E.~Evers, V.~V.~Belykh, and M.~Bayer.
Optical spin orientation of localized electrons and holes interacting with nuclei in a FA$_{0.9}$Cs$_{0.1}$PbI$_{2.8}$Br$_{0.2}$ perovskite crystal, 
\textit{ACS Photonics} \textbf{2024}, \textit{11}, 2757.

\bibitem{deQuilettes2019_si} D.~W. deQuilettes, K. Frohna, D. Emin, T. Kirchartz, V. Bulovic, D.~S. Ginger, and S.~D. Stranks, 
Charge-carrier recombination in halide perovskites.
\textit{Chem. Rev.} \textbf{2019}, \textit{119}, 11007–11019.

\bibitem{herz2017} A. D. Wright, R. L. Milot, G. E. Eperon, H. J. Snaith, M.B Johnston, and L. M. Herz,  Band-tail recombination in hybrid lead iodide perovskite. \textit{Adv. Funct. Mater.} \textbf{2017}, \textit{27}, {1700860}.

\bibitem{yugova2009} 
I. A. Yugova, M. M. Glazov, E. L. Ivchenko, A. L. Efros, 
Faraday rotation and ellipticity in an ensemble of singly charged quantum dots. 
\textit{Phys. Rev. B} \textbf{2009}, \textit{80}, 104436. 

\bibitem{zhukov2018} 
E. A. Zhukov, E. Kirstein, N. E. Kopteva, F. Heisterkamp, I. A. Yugova, V. L. Korenev, D. R. Yakovlev, A. Pawlis, M. Bayer, A. Greilich, 
Discretization of the total magnetic field by the nuclear spin bath in fluorine-doped ZnSe. 
\emph{Nat. Commun.} \textbf{2018}, \textbf{9}, 1–8. 

\bibitem{glazovbook} M. M. Glazov, \textit{Electron and Nuclear Spin Dynamics in Semiconductor Nanostructures}, Oxford University Press, UK, \textbf{2018}.

\bibitem{smirnov2020} D. S. Smirnov, E. A. Zhukov, D. R. Yakovlev, E. Kirstein, M. Bayer, and A. Greilich,
Spin polarization recovery and Hanle effect for charge carriers interacting with nuclear spins in semiconductors.
\textit{Phys. Rev. B} \textbf{2020}, \textit{102}, 235413.         
	
\bibitem{heisterkamp2015} 
F. Heisterkamp, E. A. Zhukov, A. Greilich, D. R. Yakovlev, V. L. Korenev, A. Pawlis, and M. Bayer,
Longitudinal and transverse spin dynamics of donor-bound electrons in fluorine-doped ZnSe: Spin inertia versus Hanle effect. 
\textit{Phys. Rev. B} \textbf{2015}, \textit{91}, 235432. 

\bibitem{zhukov2018si} 
E. A. Zhukov, E. Kirstein, D. S. Smirnov, D. R. Yakovlev, M. M. Glazov, D. Reuter, A. D. Wieck,
M. Bayer, and A. Greilich,
Spin inertia of resident and photoexcited carriers in singly charged quantum dots. 
\textit{Phys. Rev. B} \textbf{2018}, \textit{98}, 121304(R). 

\bibitem{Kirstein_DNSS_2023} E. Kirstein, D. S. Smirnov, E. A. Zhukov, D. R. Yakovlev, N. E. Kopteva, D. N. Dirin, O. Hordiichuk, M. V. Kovalenko, and M. Bayer.
The squeezed dark nuclear spin state in lead halide perovskites.  
\emph{Nat. Commun.} \textbf{2023}, \textit{14}, 6683.

\bibitem{Meliakov_2024PRB} S. R. Meliakov, V. V. Belykh, E. A. Zhukov, E. V. Kolobkova, M. S. Kuznetsova,
M. Bayer, and D. R. Yakovlev,
Hole spin precession and dephasing induced by nuclear hyperfine fields in CsPbBr$_3$ and CsPb(Cl,Br)$_3$ nanocrystals in a glass matrix.  
\emph{Phys. Rev. B} \textbf{2024}, \textit{110}, 235301.

\bibitem{greilich2007} A. Greilich, A. Shabaev, D. R. Yakovlev, Al. L. Efros, I. A. Yugova, D. Reuter, A. D. Wieck, and M. Bayer,
Nuclei-induced frequency focusing of electron spin coherence.
\textit{Science} \textbf{2007}, \textit{317}, 1896.

\bibitem{perez-osorio2018} 
M. A. P\'erez-Osorio, Q. Lin, R. T. Phillips, R. L. Milot, L. M. Herz, M. B. Johnston,  and F. Giustino, 
Raman spectrum of the organic–inorganic halide perovskite {CH$_3$NH$_3$PbI$_3$} from first principles and high-resolution low-temperature Raman measurement. 
\textit{J. Phys. Chem. C} \textbf{2018}, \textit{38}, 21703.

\bibitem{OptOr84} F. Meier and B. P. Zakharchenya (eds.), \textit{Optical Orientation},  Elsevier, Amsterdam, \textbf{1984}.


\end{thebibliography}

\begin{thebibliography}{60}%



\end{thebibliography}


\let\addcontentsline\oldaddcontentsline
\makeatletter
\renewcommand\tableofcontents{%
    \@starttoc{toc}%
}
\makeatother
\renewcommand{\i}{{\rm i}}


\onecolumngrid
\vspace{\columnsep}
\begin{center}
\newpage
\makeatletter
{\large\bf{Supporting Information:\\ \@title}}
\makeatother
\end{center}
\vspace{\columnsep}


\twocolumngrid

\counterwithin{figure}{section}
\renewcommand{\thepage}{S\arabic{page}}
\renewcommand{\theequation}{S\arabic{equation}}
\renewcommand{\thefigure}{S\arabic{figure}}
\renewcommand{\bibnumfmt}[1]{[S#1]}
\renewcommand{\citenumfont}[1]{S#1}

\setcounter{page}{1}
\setcounter{section}{0}
\setcounter{equation}{0}
\setcounter{figure}{0}

\section{Spin accumulation in PR curves}

\begin{figure}[ht]
 \includegraphics[width=\linewidth]{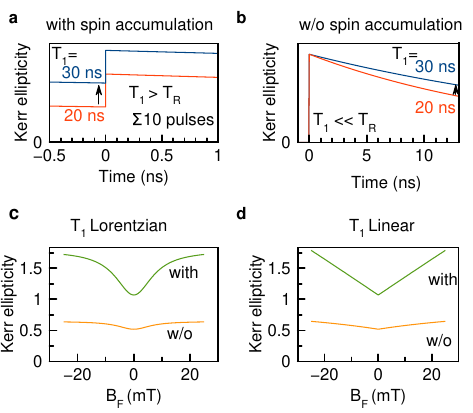}
  \caption{Spin accumulation in PR curves. The non-linear rise of the PR curves with increasing magnetic field due to the increase of $T_1$ can be motivated by the time dynamics. 
  (a) For $T_1 > T_{\rm R}$, simulated here for a sequence of 10 pulses, at small negative delay ($t^\prime = t-T_{\rm R} = -50$~ps) the rise of the KE amplitude is strongly amplified. The black arrows in ({a,b}) mark similar times of $\approx 13$\,ns.
  (b) In case of single pulse excitation, e.g. if the inverse of the laser repetition rate $1/T_{\rm R}$ is much longer than the spin dynamics time ($T_1 \ll T_{\rm R}$ ), no spin accumulation takes place and the difference in KE amplitude at $t \approx 13.2$\,ns$-50$\,ps, for either $T_1 =20$\,ns or $T_1 = 30$\,ns, is small.  (c) KE amplitude at $t\approx 13$\,ns for the cases with and without spin accumulation considered (i.e., the cases of long and short $T_{\rm R}$) as shown in ({a,b}) with Lorentzian $T_1(B) \propto 30\,\textrm{ns}-\frac{10\,\textrm{ns}}{1+(B/8\,mT)^2}$. (d) similar to (c) with linear $T_1(B) = 0.4\,\textrm{ns/mT}\cdot B+20\,\textrm{ns}$. 
	}
\label{fig:SpinAcc}
\end{figure}

To illustrate the spin accumulation effect observed in the PR curves, we calculate the spin dynamics according to $\textrm{KE}(t) = \sum_{j = 1}^{j_{\textrm{max}}} \exp{[-(t+j T_{\rm R})/T_1]}$ with $j_{\textrm{max}} = 1$ for $T_1 \ll T_{\rm R}$ and $j_\textrm{max} = 10$ for $T_1 > T_{\rm R}$, representing the cases without (w/o) and with spin accumulation. In order to be close to the experimental results, we compare cases for $T_1$ within the range of $20-30$\,ns, Figure~\ref{fig:SpinAcc}a,b. Notably, without spin accumulation the rise of amplitude for increasing $T_1$ by a factor of 1.5 from 20\,ns to 30\,ns results in an increase of the KE amplitude by a factor of 1.24, while with spin accumulation the factor is 1.67 and the overall KE amplitude is larger.

Calculations for the regime $T_1 > T_{\rm R}$ are shown in Figure~\ref{fig:SpinAcc}a, where the modeling is made for the two times $T_1=20$~ns and 30~ns. The strong spin accumulation effect is evidenced by a finite KE amplitude at negative time delays. And the effect increases for longer $T_1$ as expected. The spin accumulation is also seen through the larger amplitude at positive delays for  $T_1=30$~ns compared to the one for 20~ns. Figure~\ref{fig:SpinAcc}b shows the modeling for the regime with $T_1 \ll T_{\rm R}$, where spin accumulation is not expected. The same $T_1=20$~ns and 30~ns are taken as parameters together with $T_{\rm R}=1000$~ns.  One can see here that the KE amplitude is zero at negative delays as the spin polarization fully relaxes between the pulses. The KE signal increase at long time delays, in the comparison of the 20\,ns and 30\,ns acases, is smaller than in the case with spin accumulation.

Next we simulated different dependences of $T_1(B)$, Figure~\ref{fig:SpinAcc}c,d. With a Lorentzian dependence of $T_1$ the resulting PR curve is also Lorentzian-shaped and has the same width as the $T_1(B)$ Lorentzian. Again, one finds a much stronger contrast of the zero field to elevated field KE signal in the case where spin accumulation is considered. For the case without spin accumulation the dependence is rather flat. For a linear dependence of $T_1(B)$, in the case with spin accumulation the PR curve has a linear shape. Interestingly in the case without spin accumulation the rise of the PR curve rather follows a tangent hyperbolic shape, which might fit better to the most narrow PR peak as a Lorentzian and thus might indicate a component with rather short $T_1$ and linear $T_1(B)$. Note that in all cases the curves have been symmetrized by usage of absolute $B$ values.

\end{document}